\documentclass[12pt]{article}
\pdfoutput=1
\usepackage{putex}

\usepackage{graphicx}
\usepackage{epstopdf}
\usepackage{enumerate}
\usepackage{cite}
\usepackage{tensor}
\usepackage{slashed}

\usepackage{hyperref}

\numberwithin{equation}{section}

\newcommand{\HH}{\mathbb{H}}

\newcommand {\be} {\begin {equation}}
\newcommand {\ee} {\end {equation}}

\newcommand {\bes} {\begin {equation*}}
\newcommand {\ees} {\end {equation*}}

\newcommand{\es}[2] {\begin{equation} \label{#1} \begin{split} #2 \end{split} \end{equation}}

\newcommand{\R}{\mathbb{R}}

\newcommand{\beq}{\begin{equation}}
\newcommand{\eeq}{\end{equation}}

\begin{document}

\preprint{PUPT-2423\\MIT-CTP-4378}

\institution{IAS}{School of Natural Sciences, Institute for Advanced Study, Princeton, NJ 08540}
\institution{PU}{Department of Physics, Princeton University, Princeton, NJ 08544}
\institution{MIT}{Center for Theoretical Physics, Massachusetts Institute of Technology, Cambridge, MA 02139}

\authors{Igor R.~Klebanov,\worksat{\IAS}\footnote{On leave from
Department of Physics and Center for Theoretical Science, Princeton University.}
Tatsuma Nishioka,\worksat{\PU}  Silviu S.~Pufu,\worksat{\MIT}
Benjamin R.~Safdi\worksat{\PU}}

\date{July 2012}

\title{Is Renormalized Entanglement Entropy Stationary at RG Fixed Points?}

\abstract{
The renormalized entanglement entropy (REE) across a circle of radius $R$ has been proposed as a $c$-function in Poincar\' e invariant
$(2+1)$-dimensional field theory. A proof has been presented of its monotonic behavior as a function of $R$, based on the strong subadditivity of entanglement entropy. However, this proof does not directly establish stationarity of REE at conformal fixed points of the renormalization group.
In this note we study the REE for the free massive scalar field theory near the UV fixed point described by a massless scalar. Our numerical calculation indicates
that the REE is not stationary at the UV fixed point.
}

\maketitle

\tableofcontents

\section{Introduction}

A well-known problem in quantum field theory is how to define a  
quantity which decreases along any renormalization group (RG) trajectory. In two-dimensional QFT, an elegant solution to this problem was given by Zamolodchikov \cite{Zamolodchikov:1986gt}, who used the two-point functions of the stress-energy tensor to define a monotonic ``$c$-function.'' At RG fixed points the Zamolodchikov $c$-function equals the Weyl anomaly coefficient $c$.

An important additional property of the Zamolodchikov $c$-function is that it is stationary at the fixed points.
For example, if we consider
perturbing a CFT by a slightly relevant operator ${\cal O}$ of dimension $2-\delta $ then
\be \label{confpert}
c(g)= c_{\rm UV}- g^2 \delta  + O(g^3) \ ,
\ee
where $g$ is the renormalized dimensionless coupling. More generally, in a theory with more than one coupling constant,
\be \label{gradflow}
{\partial c\over \partial g^i}= G_{ij} \beta^j \ ,
\ee
where $G_{ij}$ is the Zamolodchikov metric, and $\beta^j= \mu {dg^j\over d\mu}$ is the beta-function for the coupling $g^j$.
The fact that the metric $G_{ij}$ is non-singular guarantees the stationarity of the Zamolodchikov $c$-function at any fixed point in two-dimensions.

It is of great interest to find out if these results extend to field theory in dimension $d>2$.
  In four-dimensional conformal field theory there are two Weyl anomaly coefficients, $a$ and $c$. Long ago Cardy conjectured \cite{Cardy:1988cwa}  that it should be the $a$-coefficient that decreases under RG flow.  This coefficient can be calculated from the expectation of value of the trace of the stress-energy tensor
in the Euclidean theory on $S^4$. Using conformal perturbation theory, it is possible to establish the analogue of (\ref{confpert}) in four-dimensions \cite{Cardy:1988cwa}:
\be 
c(g)= a_{\rm UV}- g^2 \delta  + O(g^3) \ ,
\ee
where the perturbing operator ${\cal O}$ has dimension $4-\delta$.
Over the years
considerable evidence was accumulated in favor of Cardy's conjecture, especially in supersymmetric 4-d field theories where $a$ is determined by the $U(1)_R$ charges \cite{Anselmi:1997am}. In particular, the principle of $a$-maximization \cite{Intriligator:2003jj}, which states that at superconformal fixed points the correct R-symmetry locally maximizes $a$, has passed many consistency checks that rely both on the field theoretic methods and on the AdS/CFT correspondence \cite{Maldacena:1997re,Gubser:1998bc,Witten:1998qj}.
Last year, a general proof of the $a$-theorem was constructed in \cite{Komargodski:2011vj,Komargodski:2011xv}.

A proposal for a similar theorem, called the $F$-theorem, in 3-d Euclidean CFT was made in \cite{Jafferis:2011zi}:
\es{MtheoryExpectation}{
  F = -\log |Z_{S^3}| \,,
 }
where $Z_{S^3}$ is the Euclidean path integral of the CFT conformally mapped to $S^3$.  The 3-d analogue of $a$-maximization is that the R-symmetry of
${\cal N}=2$ superconformal theories in three-dimensions locally maximizes $F$ \cite{Jafferis:2010un,Closset:2012vg}. In \cite{Jafferis:2011zi} it was further conjectured that in unitary 3-d CFT $F$ is positive, that it decreases along any RG flow, and that it is stationary at the fixed points. The three-sphere free energy is ambiguous along RG flow but is well-defined for any conformal field theory. Thus, a precise statement of the conjecture is that, if there is smooth RG flow from a UV CFT to an IR CFT, then their three-sphere free energies satisfy $F_{\rm UV} > F_{\rm IR}$. Various ${\cal N}=2$ supersymmetric examples have been presented in support of these statements; see, for example,
\cite{Jafferis:2011zi,Amariti:2011da,Klebanov:2011td}. Perturbations of a CFT on a three-sphere by slightly relevant operators were studied in \cite{Klebanov:2011gs}, and results
analogous to (\ref{confpert}) were found. In particular, the $S^3$ free energy (\ref{MtheoryExpectation}) is stationary at the fixed points.

 Closely related ideas 
 in Minkowski signature were advanced in \cite{Myers:2010tj}.  For a field theory on $\R^{2,1}$, it was proposed in \cite{Myers:2010tj} that one should consider the finite part of the entanglement entropy $S(R)$ across a circle of radius $R$. It was further shown in \cite{Casini:2011kv} (see also \cite{Dowker:2010yj}) that if the field theory in question is a CFT, this entanglement entropy agrees with the free energy of the Euclidean CFT on $S^3$:
 if such a field theory is conformal, then
\beq
S(R) = \alpha { 2\pi R \over \epsilon}  - F \,,
\label{genent}
\eeq
where $\epsilon$ is the short-distance cutoff.  

The ideas related to the quantum entanglement entropy were recently made more precise, and a proposal for a monotonic $c$-function
for a general, Poincar\'e invariant $(2+1)$-dimensional QFT has emerged. It is the so-called Renormalized Entanglement Entropy (REE) \cite{Liu:2012ee,Casini:2012ei}
\beq
{\cal F}(R)=- S(R) + R S'(R)
\ .\label{Ffcn}\eeq
An advantage of this definition is that the UV divergence proportional to the radius cancels, so that REE is manifestly finite for
theories that are conformal in the UV\@. Studies of various explicit examples, both in free field theory and in the AdS/CFT correspondence
\cite{Liu:2012ee,Klebanov:2012yf}, yielded a monotonic $c$-function ${\cal F}$. Furthermore, ${\cal F}'(R)= R S''(R)$.
It was shown in \cite{Casini:2012ei} using Strong Subadditivity of the EE, that for any Poincar\'e invariant field theory $S''(R)\leq 0$. This demonstrates that
${\cal F}(R)$ is a non-increasing function and establishes the $F$-theorem.

In this note we address a question about the proposed $c$-function ${\cal F}(R)$ that has not yet been elucidated: namely, is it
stationary for arbitrary perturbations around a CFT? In studies of holographic entanglement entropy \cite{Ryu:2006bv,Ryu:2006ef}, ${\cal F}$ turned out to be stationary \cite{Liu:2012ee,Klebanov:2012yf}, but it is not clear that this is a general property of the REE.

In a general field theory which does not have a gravity dual, the calculation of REE is a difficult problem even if we resort to
numerical methods.
In this paper we consider a particularly simple example of RG flow provided by a free massive scalar field on $\R^{2,1}$ with
the action
\be
I =-\frac 1 2 \int d^3 x\left[ (\partial_\mu \phi)^2 + m^2 \phi^2 \right]\ .
\ee
In this case there
are efficient numerical algorithms for calculating ${\cal F}$ \cite{Srednicki:1993im,Huerta:2011qi}. On dimensional grounds, ${\cal F}$  must be a function of
$g=(m R)^2$, which is the dimensionless coupling associated with the
relevant perturbation  $m^2\phi^2/2$. The UV fixed point of this theory is that of a massless free scalar, for which the value of ${\cal F}$ is known analytically \cite{Dowker:2010yj,Klebanov:2011gs}; therefore,
\es{F0}{
{\cal F}(g=0)={\cal F}_{\rm UV}={\ln 2\over 8}- {3\zeta(3)\over 16 \pi^2}\approx 0.0638\ .
}
The stationarity of the $c$-function at the UV fixed point would require ${\cal F}'(0)=0$. However, our numerical calculation
instead indicates that ${\cal F}'(g)$ is negative at small $g$. In fact, it may diverge in the limit $g\rightarrow 0$, but
the limitations of our numerical work do not allow us to make a precise statement about the small $g$ behavior of
${\cal F}'(g)$.

Our numerical results suggest that REE does not in general define a $c$-function in the Zamolodchikov sense, because it is not always stationary
at conformal fixed points.\footnote{In $(1+1)$-dimensions the monotonic $c$-function derived from the entanglement entropy of a segment
\cite{Casini:2004bw} is not stationary with respect to $m^2$ either. In $(1+1)$-dimensions this is likely due to the fact that $\phi^2$ is not a conformal primary field in the CFT of a massless scalar.} It is not clear how to compare this statement with the proof \cite{Casini:2012ei} of the monotonicity of ${\cal F}$.
If $g$ can have either sign, then ${\cal F}'(0)\neq 0$ would violate monotonicity. In the example we have considered,
however, insisting on a stable UV fixed point requires that $g$ is positive. So, there is no immediate conflict with the work of
\cite{Casini:2004bw}. We note, however, that the absence of stationarity means that an equation like (\ref{gradflow}) cannot
in general apply to ${\cal F}$.
Examination of other examples where the stationarity of ${\cal F}$ can be tested is highly desirable.

\section{Strip entanglement entropy at small mass}
Although the renormalized entanglement entropy across a circle is of our main interest, as a warm-up we start with a slightly different but simpler quantity, the entanglement entropy of a strip.

The entanglement entropy between the strip of width $R$ and its complement in $(2+1)$-dimensional free field theory is simply related to the entanglement entropy between the interval of length $R$ and its complement in the $(1+1)$-dimensional theory.  More precisely, for the $(2+1)$-dimensional massive scalar field of mass $m$ we have the relation~\cite{Casini:2005zv}
\es{Sstrip}{
S^{(2+1)}_{\text{strip}}(m, R) = {L \over \pi} \int_0^\infty dp \, S^{(1+1)}_{\text{interval}}\big(\sqrt{p^2 + m^2}, R \big) \,,
}
where $\sqrt{p^2 + m^2}$ is the mass of the $(1+1)$-dimensional scalar field, and $L$ is the length of the strip.  To derive this equation one first compactifies the direction parallel to the strip to a large circle of length $L$.  Decomposing the $(2+1)$-dimensional scalar field into angular momentum modes along this circle and then taking $L \to \infty$ leads to~\eqref{Sstrip}.

 It is useful to define the entropic $c$-function for the strip through
 \be \label{stripc}
 C_{\text{strip}}  \equiv R^2 \, \partial_R \, \hat S^{(2+1)}_{\text{strip}}\ ,
  \ee
  where $\hat S^{(2+1)}_{\text{strip}} \equiv S^{(2+1)}_{\text{strip}} / L$ is the entanglement entropy per unit length. This function is manifestly finite and cut-off independent \cite{Nishioka:2006gr,Klebanov:2007ws}.   For the massive scalar field, $ C_{\text{strip}}$ is also simply related to the $(1+1)$-dimensional entropic $c$-function, defined \cite{Casini:2004bw} by $c(t=mR) \equiv R \, \partial_R \, S^{(1+1)}_{\text{interval}}$, through~\eqref{Sstrip}:
 \es{Candc}{
C_{\text{strip}}(mR) = {1 \over 2\pi} \int_{-\infty}^\infty dx\ c(\sqrt{x^2 + (mR)^2}  ) \,.
}
We should note, however, that in general $ C_{\text{strip}}$ is not expected to be a good $c$-function; for example, it is not constant
 along lines of fixed points.

\begin{figure}[htb]
  \leavevmode
\begin{center}
\scalebox{1}{\includegraphics{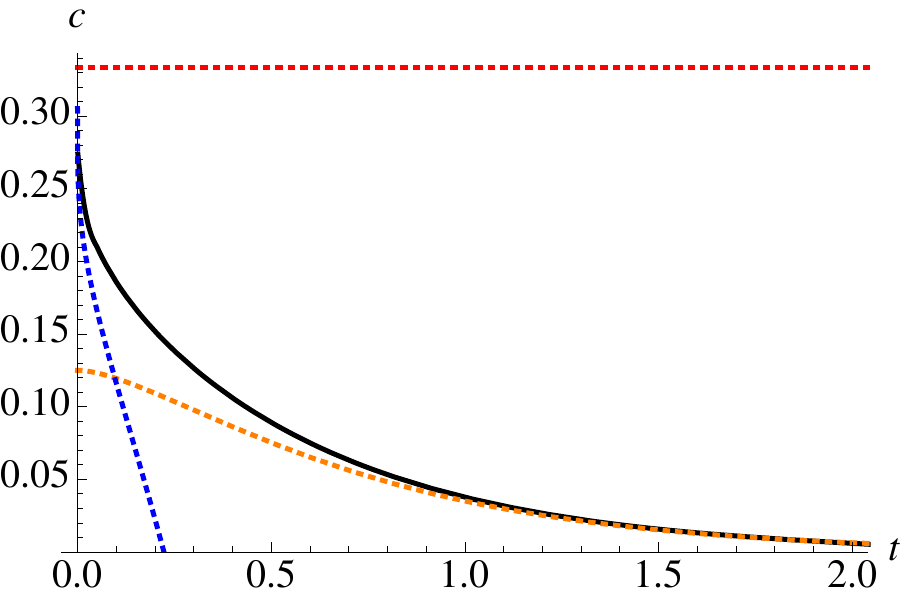}}
\end{center}
\caption{The entropic $c$-function $c \equiv R \, \partial_R \, S^{(1+1)}_{\text{interval}}$ for the $(1+1)$-dimensional massive scalar field as a function of $t \equiv m R$, where $m$ is the mass and $R$ is the length of the interval.  The black curve comes from a numerical calculation using the prescription in~\cite{Casini:2005zv}.  The blue and orange curves are the analytic approximations in~\eqref{smallLarge} at small and large values of $t$, respectively.  The dotted red line marks the conformal value $c(0) = 1/3$.     }
\label{S2Plot}
\end{figure}

The entropic $c$-function for the $(1+1)$-dimensional massive scalar field is a well studied quantity (see~\cite{Casini:2009sr} for a review).  The function interpolates between $1/3$ in the UV ($t = 0$) and $0$ in the IR ($t \to \infty$).
At small and large values of $t$ one can calculate the leading behavior of $c(t)$ analytically~\cite{Casini:2005zv},
\es{smallLarge}{
c(t) &= {1 \over 3} + {1 \over 2 \log t} + \cdots \,, \qquad t \ll 1 \,, \\
c(t) &= {t \over 4} K_1 (2 t) + \cdots \,,  \qquad t \gg 1 \,.
}
Unfortunately there is no known closed form expression for $c(t)$ along the entire RG flow.  The function may be constructed numerically by solving an infinite sequence of nonlinear differential equations~\cite{Casini:2005zv}.  Following this procedure leads to the curve in figure~\ref{S2Plot}.

\begin{figure}[htb]
  \leavevmode
\begin{center}
\scalebox{1}{\includegraphics{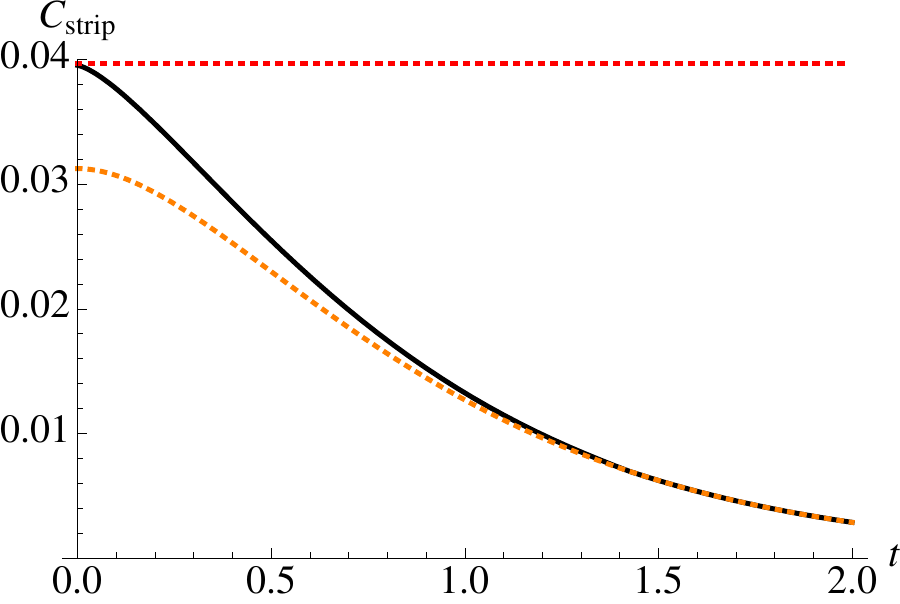}}
\end{center}
\caption{The function $ C_{\text{strip}}  \equiv R^2 \, \partial_R \, \hat S^{(2+1)}_{\text{strip}}$ for the free massive scalar field in black, where $\hat S^{(2+1)}_{\text{strip}} \equiv S^{(2+1)}_{\text{strip}} / L$ is the entanglement entropy per unit length across the strip of width $R$.  The orange curve is the IR approximation in~\eqref{IRstrip}.  The dotted red line is the initial value at $t = mR = 0$ given in~\eqref{C0}.      }
\label{Cstrip}
\end{figure}

The function $ C_{\text{strip}} (mR)$ is a monotonically decreasing function that approaches zero exponentially fast in the IR\@.  Using~\eqref{Candc} one may determine numerically its value at $mR = 0$~\cite{Casini:2005zv},
 \es{C0}{
 C_{\text{strip}} ( 0) = {1 \over \pi}  \int_0^\infty dt \, c( t ) \approx 3.97 \times 10^{-2} \,.
}
At large $mR$ we may use the approximation for $c(t)$ in the second line of~\eqref{smallLarge} to write
\es{IRstrip}{
C_{\text{strip}} (mR) \approx {1 \over 16} e^{-2 mR} \left( mR + {1 \over 2} \right) \,, \qquad mR \gg 1 \,.
}
In figure~\ref{Cstrip} we numerically plot $C_{\text{strip}}$ along the RG flow together with the IR analytic approximation.

\begin{figure}[htb]
\leavevmode
\begin{center}
\scalebox{1}{\includegraphics{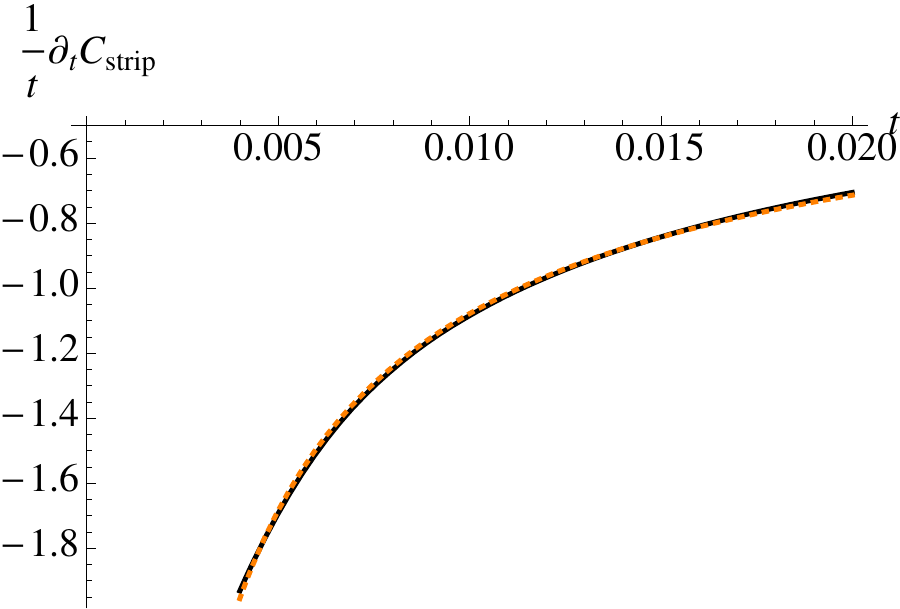}}
\end{center}
\caption{The first derivative $ C_{\text{strip}}'(t)/t$ as $t = mR \to 0$.  The black curve comes from the numerical calculation, and the dotted orange curve comes from fitting the numerics to the function $C_{\text{strip}}'(t)/t \approx a/(t \log^2 t) $ as $t \to 0$.  We find $a \approx -0.25$, which agrees with the analytic result in~\eqref{Cstder}.  This
means that $C_{\text{strip}}$ is not stationary for the massive scalar field.  }
\label{secondD}
\end{figure}

It is interesting to ask whether $C_{\text{strip}}$ is stationary at $mR = 0$.
The first derivative of the entropic $c$-function \eqref{Candc} with respect to $t=mR$
gives
\es{Cstder}{
	C'_\text{strip}(t) &= \frac{1}{2\pi} \int_{-\infty}^\infty dx \frac{t}{\sqrt{x^2 + t^2}}\, c'(\sqrt{x^2 + t^2})  \\
	&=  \frac{1}{2\pi} \int_{-\infty}^\infty dx \frac{1}{\sqrt{x^2 + 1}} \left[ -\frac{1}{2\sqrt{x^2 + 1} \log^2 (t\sqrt{x^2 + 1})} + O\left(  \log^{-3} t \right) \right]  \\
	&\xrightarrow[t \to 0]{} -\frac{1}{4\log^2 t} \, ,
}
where we used the asymptotic form of the $(1+1)$-dimensional entropic $c$-function \eqref{smallLarge} and rescaled $x\to t x $ in the second line.  While this implies that  $C_{\text{strip}}'(0) = 0$, stationarity additionally requires that
$C'_\text{strip}(t) / t$ vanishes at $t = 0$.  This is clearly not the case since this quantity
diverges as $-1 / (4 t \log^2 t)$ as $t \to 0$.
As can be seen in figure~\ref{secondD}, we confirm this behavior numerically by plotting $C'_\text{strip}(t) / t$ along with the analytic prediction.

To summarize the results of this section, we have found that the small mass expansion of the UV finite function (\ref{stripc}) for a strip of width $R$ is
\be
C_{\text{strip}}(mR) \approx 0.0397- {|mR|\over 4\log^2 (mR)} + \cdots\ .
\ee
In the next section we will present evidence that the REE across a circle of radius $R$ has a similar structure, with $d{\cal F}/dg$
looking clearly negative at small $g=(mR)^2$.

\section{Disk entanglement entropy at small mass}

The entanglement entropy for the massive scalar field across the circle of radius $R$ may be calculated numerically following the prescription in~\cite{Huerta:2011qi,Casini:2009sr,Srednicki:1993im,2003JPhA...36L.205P}.  This numerical method has passed many non-trivial checks.  For example, in~\cite{Huerta:2011qi} the terms proportional to $(mR)$ and $1 / (mR)$ in the large $(mR)$ expansion of the disk entanglement entropy were matched numerically to the analytic predictions to high accuracy.  A similar analysis in~\cite{Safdi:2012sn} matched the $1 / (mR)^3$ term in the large mass expansion of the EE to the analytic prediction, and in~\cite{Liu:2012ee} the value of the REE at $mR = 0$ was shown to agree with~\eqref{F0}.

The numerical procedure~\cite{Huerta:2011qi,Casini:2009sr,Srednicki:1993im,2003JPhA...36L.205P} expands the scalar field into modes of integer angular momentum $n$.  The radial direction is discretized into a lattice of $N$ units.  For each $n$ the discrete Hamiltonian takes the form $H_n = {1\over2} \sum_i \pi_i^2 + {1\over2} \sum_{ij} \phi_i K_n^{ij} \phi_j$, where $\pi_i$ is the conjugate momentum to $\phi_i$ and $i = 1, \dots, N$.  The nonzero entries of the  $N \times N$ matrix $K_n$ are
\es{Kmat}{
K_n^{11} = {3 \over 2} + n^2 + m^2 \,, \qquad K_n^{ii} = 2 + {n^2 \over i^2} + m^2 \,, \qquad K_n^{i,i+1} = K_n^{i+1,i} = - {i+1/2 \over \sqrt{i(i+1)} } \,.
}

To compute the entanglement entropy we need to know the two-point correlators $X \equiv \left< \phi_i \phi_j \right> ={1\over 2} (K^{-1/2})_{ij}$ and $P \equiv \left< \pi_i \pi_j \right> ={1\over 2} (K^{+1/2})_{ij}$.  If the radius of the entangling circle $R$ is a half-integer in units of the lattice spacing, then we must reduce the matrices $X_{ij}$ and $P_{ij}$ to the $r \times r$ matrices $X^r_{ij}$ and $P^r_{ij}$, which are defined by taking  $1 \leq i,j \leq r$ with $r = R - {1 \over 2}$.  The entanglement entropy is then given by
\es{NumS}{
S(R) = S_0 + 2 \sum_{n=1}^\infty S_n \,,
}
with
\es{Sn}{
S_n = \tr \left[ \left( \sqrt{X_n^r P_n^r} + \frac12 \right) \log \left( \sqrt{X_n^r P_n^r }+ \frac12 \right) - \left( \sqrt{X_n^r P_n^r} - \frac12 \right) \log \left( \sqrt{X_n^r P_n^r} - \frac12 \right) \right] \,.
}
Further details of the numerical calculation can be found in~\cite{Huerta:2011qi}.

In order to achieve the continuum limit we need to take $N \gg r \gg 1$ and $m\ll 1$ (restoring the lattice spacing, the latter condition is $m\epsilon \ll 1$). Then $mR$ can be either small or large, and we can explore the REE as a function of this parameter.

In our calculations we use a radial lattice consisting of $N = 200$ points.  We want to calculate the entanglement entropy for $.06 < mR < 2$, but to minimize lattice effects we restrict $30 < r < 50$ in lattice units.  To accomplish this we calculate the entanglement entropy for $ m = .002\cdot i$ in inverse lattice units, with $i = 1, \dots, 20$.
To take into account finite lattice effects, we follow~\cite{Liu:2012ee} and for the lowest $10$ angular momentum modes we repeat the calculation on lattices of sizes $N = 200+10\cdot j$, with $j = 0, \dots,50$, and then extrapolate to $N = \infty$.
We perform the numerical calculation for the first $3000$ angular momentum modes.  We take into account higher angular momentum modes by
using the asymptotic behavior of $S_n$ derived in Appendix~\ref{LARGEN}.

\begin{figure}[htb]
  \leavevmode
\begin{center}
\scalebox{1}{\includegraphics{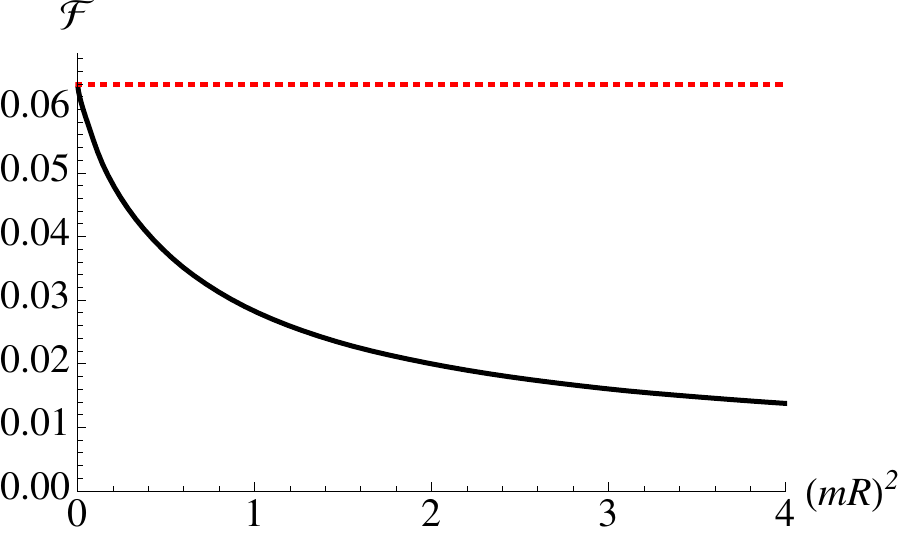}}
\end{center}
\caption{The renormalized entanglement entropy ${\cal F}$ across the circle of radius $R$ for the massive real free scalar plotted in black as a function of $(mR)^2$.
In this plot it can clearly be seen that $\partial_{(mR)^2} {\cal F}$ is negative and nonzero at $(mR)^2 = 0$, which implies that the REE ${\cal F}$ is not stationary at the UV fixed point of a free massless scalar field.
The dotted red line is the zero mass value ${\cal F}_{\rm UV}={\ln 2\over 8}- {3\zeta(3)\over 16 \pi^2}$.}
\label{scalarPlot}
\end{figure}

From the entanglement entropy we construct the renormalized entanglement entropy ${\cal F}$ along the RG flow.  A plot of this function versus $(mR)^2$ is given in figure~\ref{scalarPlot}.
From this plot we can see that $\partial_{(mR)^2} {\cal F}$ is clearly negative and nonzero at $(mR)^2 = 0$.  Moreover, it appears that this quantity might actually diverge as $(mR)^2 \to 0$, though our limited numerical precision prevents us from making any precise statements along these lines.

\section{A toy model: the massive scalar on $\HH^2\times S^1$}

For a CFT in $(2+1)$-dimensions the calculation of the R\' enyi $q$-entropy of a disk may be mapped to a calculation on $\HH^2\times S^1$
\cite{Casini:2010kt,Casini:2011kv},
where $q$ is the radius of the circle. For a free scalar field of mass $m$ the free energy on this space is \cite{Klebanov:2011uf}
\es{FqOneInt}{
  {\cal F}_q(m^2) = \int_0^\infty d \lambda {\cal D}(\lambda)
   \left[ \log \left(1 - e^{-2 \pi q \sqrt{\lambda + (mR)^2 } }  \right)
    + \pi q  \sqrt{\lambda + (mR)^2 }\right] \,,
 }
 where the density of states is given by
 \es{DLambda}{
  {\cal D}(\lambda) d\lambda = \frac{\Vol(\HH^2)}{4 \pi} \tanh (\pi \sqrt{\lambda}) d\lambda \ ,
 }
 and the regularized volume of the hyperbolic space is $\Vol(\HH^2)=-2\pi$.
 The formula for the finite part of entanglement entropy is
 \be
 S_1= {\partial {\cal F}_q\over \partial q}\bigg|_{q=1}-  {\cal F}_1
 \ .
 \ee
 Applying this relation at $m=0$, which corresponds to a massless scalar on a disk, one obtains
 $S_1= -{\ln 2\over 8}+ {3\zeta(3)\over 16 \pi^2}$, in agreement with other methods \cite{Klebanov:2011uf}.

 The status of the calculation on $\HH^2\times S^1$ for $m^2>0$ is less clear since this does not correspond to turning on mass
 on the original disk geometry. It is still interesting to inquire whether $S_1$ defined above is stationary with respect to turning on
 $m^2$. An explicit calculation yields
\be\label{EEmass}
 {\partial S_1\over \partial m^2}\bigg|_{m^2 = 0}= {\pi^2\over 16} R^2\,.
 \ee
 Similarly, the calculation of the R\' enyi entropy can be mapped to that on the $q$-fold
covering of $S^3$ \cite{Klebanov:2011uf}. For a massive free scalar field of mass $m$ on
this space, the derivative of $S_1$ with
respect to the mass of the scalar field is
$ {\partial S_1\over \partial m^2}|_{m^2 = 0} = -{\pi^2\over 16} R^2$, which agrees in absolute value with (4.4) but has a different sign. 
Either way,
$S_1$ is not stationary in these toy models.
 This lack of stationarity of $S_1$, which is easily established analytically, is reminiscent of the numerical result for the disk entanglement
 found in the previous section.

\section*{Acknowledgments}
We thank H.~Casini, Z.~Komargodski, D.~Kutasov, S.~Sachdev and E.~Witten
 for helpful discussions.
 The work of IRK and TN was supported in part by the US NSF under Grant No.~PHY-0756966. IRK gratefully acknowledges support from the IBM Einstein Fellowship at the Institute for Advanced Study, and from the John Simon Guggenheim Memorial Fellowship.  SSP was supported by a Pappalardo Fellowship in Physics at MIT and by the U.S. Department of Energy under cooperative research agreement Contract Number DE-FG02-05ER41360\@.  BRS was supported by the NSF Graduate Research Fellowship Program. IRK is grateful to the Aspen Center for Physics and the NSF Grant No. 1066293 for hospitality during the completion of this paper. BRS thanks the Institute for Advanced Study for hospitality.

\appendix

\section{Large $n$ asymptotics of $S_n$}
\label{LARGEN}

A large $n$ expansion of $S_n$ is made possible by the fact that to leading order in $n$ the matrix $K_n^{ij}$ defined in \eqref{Kmat} is diagonal.  Therefore, to leading order in $n$, the matrices $X_n = \frac 12 K_n^{-1/2}$ and $P_n = \frac 12 K_n^{1/2}$ are also diagonal, and so are their truncated counterparts $X_n^r$ and $P_n^r$.  Consequently, in this approximation $X_n^r P_n^r = \frac 14$, and from \eqref{Sn} we conclude that $S_n$ vanishes.  The non-trivial dependence of $S_n$ on $n$ comes from the $1/n$ corrections to the relation $X_n^r P_n^r = 1/4$, which we will now calculate.

From $P_n^2 = \frac 14 K_n$, we find for $i>2$
 \es{PExpansion}{
  P_n^{ii} &= \frac{n}{2 i} + \frac{i (m^2 + 2)}{4 n} - \frac{i^3 (m^4 + 4 m^2 + 6)}{16 n^3} + O(n^{-5}) \,, \\
  P_n^{i, i+1} &= P_n^{i+1, i} = - \frac{\sqrt{i(i+1)}}{4n} + \frac{i^{3/2} (i+1)^{3/2} (2 + m^2)}{8 n^3} + O(n^{-5}) \,, \\
  P_n^{i, i+2} &= P_n^{i+2, i} = -\frac{i^{3/2} (i+2)^{3/2}}{16 n^3} + O(n^{-5}) \,.
 }
In general, $P_n^{i, i+a} = O(1/n^{2a-1})$, so up to a fixed order in $1/n$ the matrix $P_n$ is almost diagonal.  From the relation $X_n P_n = \frac 14$, we can obtain a similar expansion for the entries of $X_n$:
 \es{XExpansion}{
  X_n^{ii} &= \frac{i}{2 n} - \frac{i^3 (m^2 + 2)}{4 n^3} + \frac{i^3 \left(4 + 3 i^2 \left( m^4 + 4 m^2 + 6 \right) \right)}{16 n^5} + O(n^{-7}) \,, \\
  X_n^{i, i+1} &= X_n^{i+1, i} = \frac{i^{3/2} (i+1)^{3/2}}{4n^3}
    - \frac{i^{3/2} (i+1)^{3/2} \left(3i^2 + 3i + 1 \right)(m^2 + 2)}{8 n^5} + O(n^{-7}) \,, \\
  X_n^{i, i+2} &= X_n^{i+2, i} = \frac{i^{3/2} (i+2)^{3/2} \left(3 i^2 + 6i + 2 \right)}{16n^5} + O(n^{-7}) \,,
 }
with the general asymptotic form $X_n^{i, i+a} = O(1/n^{2a + 1})$.  By definition, $X_n P_n = 1/4$, or explicitly
 \es{XPExplicit}{
  (X_n P_n)^{i, i+a} = \sum_b X_n^{i, i+b} P_n^{i+b, i+a} = \frac{\delta_{a0}}{4} \,.
 }
The product of the reduced matrices $X_n^r$ and $P_n^r$, however, fails to equal $1/4$ precisely because some of the terms in the sum over $b$ are absent, namely the ones with $i+b >r$.  Using the asymptotic expansions \eqref{PExpansion}--\eqref{XExpansion}, we find that up to $O(n^{-8})$ corrections, the matrix $X_n^r P_n^r - \frac 14$ is identically zero except for its lower-right two-by-two block:
 \es{TwoByTwo}{
  (X_n^r P_n^r )^{rr} &= \frac 14 + \frac{r^2 (r+1)^2}{16 n^4} - \frac{(m^2 + 2) r^2 (r+1)^2 (2r + 1)^2}{32 n^6} + O(n^{-8}) \,, \\
  (X_n^r P_n^r)^{r, r-1} &= \frac{r^{3/2} (r-1)^{3/2} (r+1)^3}{64 n^6} + O(n^{-8}) \,, \\
   (X_n^r P_n^r)^{r-1, r} &= \frac{r^{1/2} (r-1)^{3/2} (r+1)^2 (3r^2 - 1)}{64 n^6} + O(n^{-8}) \,,
 }
and $(X_n^r P_n^r)^{ij} = \delta_{ij}/4 + O(n^{-8})$ for all other $1 \leq i, j \leq r$.

A straightforward computation that focuses on the lower-right two-by-two block of $X_n^r P_n^r$ shows that the matrix $\sqrt{X_n^r P_n^r} + \frac 12$ that appears in the expression for $S_n$ \eqref{Sn} has $r-1$ eigenvalues equal to $1 + O(n^{-8})$ and one eigenvalue equal to
 \es{EvaluePlus}{
  1 + c_n - c_n \frac{(m^2 + 2)(2r+1)^2}{2n^2} + O(n^{-8}) \,, \qquad
   c_n \equiv \frac{r^2 (r+1)^2}{16 n^4} \,.
 }
Similarly $\sqrt{X_n^r P_n^r} - \frac 12$ has $r-1$ eigenvalues equal to $O(n^{-8})$ and one eigenvalue equal to
 \es{EvalueMinus}{
  c_n - c_n \frac{(m^2 + 2)(2r+1)^2}{2n^2} + O(n^{-8}) \,.
 }
Substituting these results into the expression for $S_n$ in \eqref{Sn} we obtain
 \es{GotSnAsymp}{
  S_n = c_n \left( 1 - \log c_n \right) + c_n \log c_n
     \frac{(m^2 + 2)(2r+1)^2}{2 n^2} + O(\log n / n^8) \,.
 }

\bibliographystyle{ssg}
\bibliography{CGLP}

\end{document}